\documentclass[conference, a4paper]{IEEEtran}

\usepackage[T1]{fontenc}% optional T1 font encoding
\usepackage{cite}

\ifCLASSINFOpdf
  \usepackage[pdftex]{graphicx}
  \graphicspath{{./pdf/}}
  \DeclareGraphicsExtensions{.pdf}
\else
  \usepackage[dvips]{graphicx}
  \graphicspath{{./eps/}}
  \DeclareGraphicsExtensions{.eps}
\fi

\usepackage{amsmath}
\usepackage{mathdots}
\usepackage{mathtools}
\usepackage[cmintegrals]{newtxmath}
\usepackage{bm}
\usepackage{balance}
\usepackage{url}

\usepackage[utf8]{inputenc}

\DeclareMathOperator*{\argmin}{arg\,min}

\begin{document}
\title{Prototyping Software Transceiver for the 5G New Radio Physical Uplink Shared Channel}

\author{\IEEEauthorblockN{Grzegorz Cisek and Tomasz P. Zieliński}
\IEEEauthorblockA{AGH University of Science and Technology, Department of Telecommunications, 30-059 Cracow, Poland}
\textit{gcisek@agh.edu.pl}}

\maketitle

\begin{abstract}
5G New Radio (NR) is an emerging radio access technology, which is planned to succeed 4G Long Term Evolution (LTE) as global standard of cellular communications in the upcoming years. This paper considers a digital signal processing model and a software implementation of a complete transceiver chain of the Physical Uplink Shared Channel (PUSCH) defined by the version 15 of the 3GPP standard, consisting of both baseband transmitter and receiver chains on a physical layer level. The BLER performance of the prototype system implementation under AWGN and Rayleigh fading channel conditions is evaluated. Moreover, the source code of high-level numerical model was made available online on a public repository by the authors. In the paper's tutorial part, the aspects of the 5G NR standard are reviewed and their impact on different functional building blocks of the system is discussed, including synchronization, channel estimation, equalization, soft-bit demodulation and LDPC encoding/decoding. A review of State-of-Art algorithms that can be utilized to increase the performance of the system is provided together with a guidelines for practical implementations.
\end{abstract}

\begin{IEEEkeywords}
5G; New Radio; PUSCH; SDR; OFDM; channel estimation; LDPC.
\end{IEEEkeywords}

\section{Introduction}

A growing demand for more advanced wireless data transmission services has always been the driver for the development of radio access technology standards. The 5G New Radio (NR) being currently standardized by the 3GPP is designed to extend the 4G LTE spectral efficiency, latency and data rate, as well as expanding the current frequency coverage to millimeter wave (mmWave) frequencies up to 100 GHz. The standardization is focused on providing a support for diversified use cases, including Ultra-Reliable Low Latency Communications (URLLC), Internet of Things (IoT) and Massive Machine Type Communications (mMTC). The worldwide commercial launch of 5G NR is planned for the year 2020.

The Physical Uplink Shared Channel (PUSCH) in 5G NR is designated to carry multiplexed control information and user application data. The structure of the channel specified by the 5G NR standard is a modification of PUSCH channel defined by the 4G LTE standard \cite{Hou13}. The 5G NR provides much more flexibility and reliability comparing to its predecessor, including more elastic pilot arrangements and support for both CP-OFDM and DFT-s-OFDM waveforms \cite{Levanen18}. The standard introduced filtered OFDM (f-OFDM) technique, which adds additional filtering to reduce Out-of-Band emission and improve the performance at higher modulation orders \cite{DiStasio18}. Moreover, modifications in Forward Error Correction (FEC) were imposed to replace the Turbo Codes used in 4G LTE by Quasi-Cyclic Low Density Parity Check (QC-LDPC) codes \cite{Richardson18}, which were proven to achieve better transmission rates and provide opportunities for more efficient hardware implementations \cite{Shao19}.

In this paper, a prototype software implementation of the PUSCH channel is provided. The transmitter and the receiver are based on 3GPP 5G New Radio (NR) physical layer specification and implement end-to-end PUSCH channel structure as defined by \cite{TS38211, TS38212}. The model considers various receiver algorithms that are used for synchronization, channel deconvolution and symbol detection. The selected algorithms are well founded in the literature and have been successfully used in implementations of the legacy 3G and 4G receivers. It is demonstrated that the same algorithms may be successfully used in implementation of the 5G NR communication system with minor modifications.

\section{Aspects of the 5G NR Standard}

\subsection{Frame Structure}

\begin{figure}[!t]
  \centering
  \includegraphics[width=3.2in]{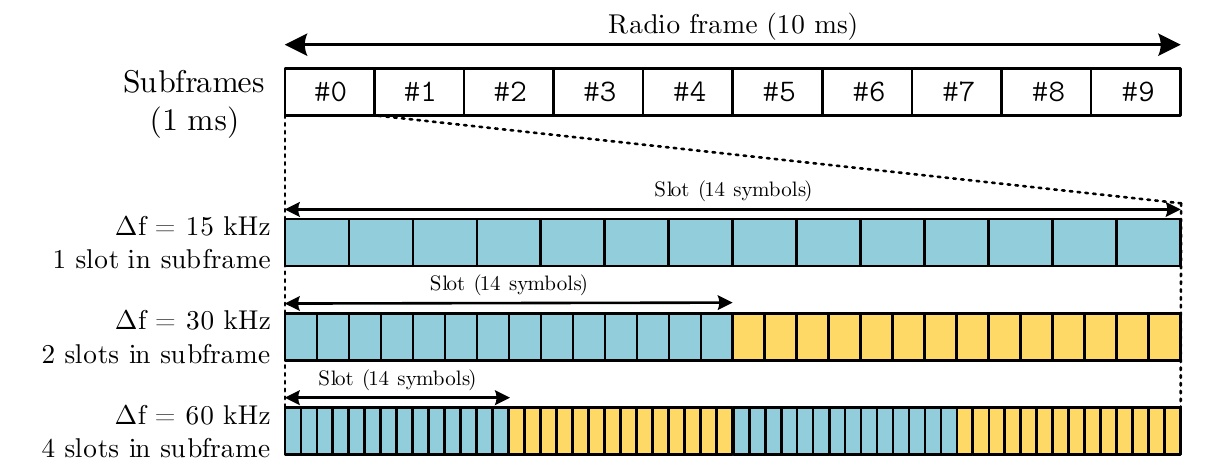}
  \caption{Scalable frame structure in 5G NR.}
  \label{fig_frame_struct}
\end{figure}

Similarly as in case of LTE, transmission of 5G NR downlink and uplink data is organized into frames of 10 ms duration, each divided into 10 subframes of 1 ms each. Subframes are composed of a variable number of slots, depending on selected subcarrier spacing \(\Delta f\), which is parameterized in 5G NR. The presence of multiple OFDM numerologies is a new concept comparing to 4G LTE, introduced to provide more flexibility in term of cell size, latency and interference resilience \cite{Zaidi18}. A slot is built from 14 OFDMA symbols, each prepended with a cyclic prefix. A subcarrier that is located within a passband and is designated for transmission is called a Resource Element (RE) - the smallest unit of the resource grid. A group of 12 neighboring RE in the same symbol form a Physical Resource Block (PRB). Flexible frame structure of 5G NR standard has been shown in Fig. \ref{fig_frame_struct} and the OFDM numerology scaling has been summarized in Table \ref{table_numerology}.

\subsection{Pilot Structure}

The 5G NR standard defined two types of reference signals associated with transmission of the PUSCH channel.

Demodulation Reference Signal (DMRS) is user specific reference signal with high frequency density. The DMRS is transmitted within dedicated OFDMA symbols only and designated for frequency-selective channel estimation. A number of DMRS symbols within a slot may vary between 1 and 4 depending on configuration, when a denser DMRS symbol spacing in time is designated for fast time-varying channels to obtain more accurate estimates within the coherence time of the channel \cite{Noh17}. In frequency domain, DMRS PRB are mapped within the whole transmission allocation. A spacing between DMRS RE assigned for the same Antenna Port (AP) may be chosen between 2 and 3. In case of \(2 \times 2\) MIMO, the standard allows for orthogonal assignment of RE between AP. Consequently, the receiver may perform partial SIMO channel estimation based on the DMRS RE prior to MIMO equalization, neglecting spatial correlation.

The second type of the reference signal is Phase Tracking Reference Signal (PTRS) \cite{Qi18}. The PTRS subcarriers are arranged in a comb structure having high density in time domain. It is used mainly in mmWave frequency bands to track and correct the phase noise, which is a considerable source of performance losses. Usage of the PTRS is optional, as it may lower a total spectral efficiency of the transmission when the effect of phase noise is negligible.

\setlength\tabcolsep{5pt}
\begin{table}[!t]
\renewcommand{\arraystretch}{1.2}
\caption{Scalable OFDM numerology in 5G NR.}
\label{table_numerology}
\centering
\begin{tabular}{ccccc}
\hline
Subcarrier Spacing \(\Delta f\) [kHz]  & 15     & 30     & 60     & \(15 \cdot 2^n\)\\
\hline
OFDM Symbol duration [\(\mu s\)]       & 66.67  & 33.33  & 16.67  & \(66.67 / 2^n\)\\
Cyclic prefix duration [\(\mu s\)]     & 4.69   & 2.34   & 1.17   & \(4.69 / 2^n\) \\
Slot duration (14 symbols) [\(\mu s\)] & 1000   & 500    & 17.84  & \(1000 / 2^n\) \\
Slots in subframe                      & 1      & 2      & 4      & \(2^n\) \\
Sampling rate* [MHz]                   & 30.72  & 61.44  & 122.88 & \(30.72 \cdot 2^n\) \\
\hline
& & & & \\[-3mm]
\end{tabular}
\raggedright 
\scriptsize (*) assuming 2048 OFDM tones per symbol.
\end{table}

\subsection{Channel Coding}

LDPC code is a linear block code defined by a nullspace of a sparse parity check matrix \(\bm{H}\), so that for each valid codeword \(\bm{d}\) the parity check equation \(\bm{H} \bm{d} = \bm{0}\) is satisfied. The LDPC code subclass used in 5G NR standard, Quasi-Cyclic LDPC (QC-LDPC), constructs parity check matrices from multiple square sub-matrices, where each can be either a zero matrix or a circulant permutation matrix. In 5G NR, the LDPC code is defined by a lifting size \(Z_c\), determining the the sub-matrix size, and the Base Graph (BG) number. The standard defines two Base Graphs, BG1 and BG2, offering a trade-off between codeword length and achievable code rate. Each BG determines the structure of the matrix \(\bm{H}\) through a set of circular shift coefficients \(V_{i,j}\) that are defined for each individual permutation sub-matrix.

The encoding and decoding chains in 5G NR defined by \cite{TS38212} have similar structure as the ones used in Turbo code based LTE encoders. The main building blocks of the 5G NR encoder are: codeblock segmentation, CRC attachment, channel coding and rate matching with bit interleaving. The rate matching in the receiver allows for utilization of Incremental Redundancy Hybrid Automatic Repeat Request (IR-HARQ) mechanism, which is implemented by soft-combining of demodulated bits that are accumulated from multiple retransmissions of a given codeblock. A detailed analysis of rate matching design can be found in \cite{Hamidi18}.

\section{Transceiver Implementation}

\begin{figure*}[!t]
  \centering
  \includegraphics[width=0.95\textwidth]{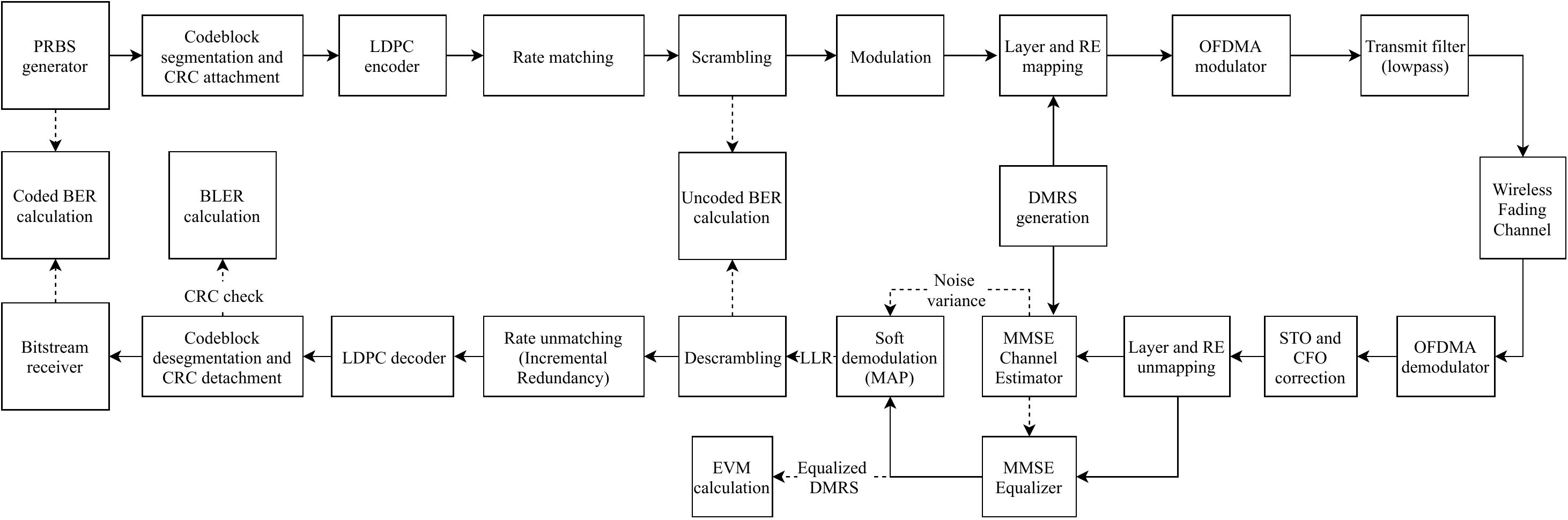}
  \caption{Transmitter and receiver chain of the 5G NR PUSCH channel.}
  \label{fig_diagram}
\end{figure*}

The transceiver implementation supports multiplexing of multiple User Equipment (UE) with \(2 \times 2\) MIMO Spatial Multiplexing (SM). The end-to-end data-flow, including transmitter and receiver chains, is shown in Fig. \ref{fig_diagram}. 

\subsection{LDPC Encoder}

The key building block of the transmitter that heavily impacts the overall complexity and performance is LDPC encoder. For the sake of the transmitter implementation, the LDPC encoder proposed by Richardson and Urbanke was chosen \cite{Richardson01}. The structure of LDPC parity check matrix used in 5G NR standard allows for further simplification of the original encoder structure.

The encoder structure in \cite{Richardson01} is based on partitioning of the parity check matrix \(\bm{H}\) of size \(m \times n\), where \(n\) is a codeword length and \(m\) is a number of redundancy bits, into the submatrices:
\begin{equation}
\label{eqn_ldpc_parity}
\bm{H} = 
 \begin{bmatrix}
  \bm{C} & \bm{D} & \bm{E} \\
  \bm{A} & \bm{B} & \bm{T} 
 \end{bmatrix}
\end{equation}
where \(\bm{T}\) is an upper triangular matrix of size \((m-g) \times (m-g)\), \(\bm{A}\) is \((m-g) \times (n-m)\), \(\bm{B}\) is \((m-g) \times g\),
\(\bm{C}\) is \(g \times (n-m)\), \(\bm{D}\) is \(g \times g\) and \(\bm{E}\) is \(g \times (m-g)\). Taking into assumption the structure of 5G NR LDPC parity check matrices based on both BG1 and BG2 defined in \cite{TS38212}, the following conditions are true for all configurations:
\begin{itemize}
\item \(g = 4 \cdot Z_c\) for both BG1 and BG2,
\item \(\bm{E} = \bm{0}\) (zero matrix),
\item \(\bm{T} = \bm{I}\) (identity matrix),
\item \(\bm{D}\) is non-singular.
\end{itemize}

Because the code is systematic, a codeword vector is given by \(\bm{d} = [\bm{s}, \, \bm{p_1}, \, \bm{p_2}]^T\), where \(\bm{s}\) is a vector of information bits and \(\bm{p_1}\), \(\bm{p_2}\) are vectors of parity bits. Taking into account the presented assumptions, a simplified version of LDPC encoder from \cite{Richardson01} dedicated for 5G NR can be formulated as:
\begin{equation}
\label{eqn_ldpc_enc_p1}
\bm{p_1}^T = \bm{D}^{-1} \bm{C} \bm{s}^T
\end{equation}
\begin{equation}
\label{eqn_ldpc_enc_p2}
\bm{p_2}^T = \bm{A} \bm{s}^T + \bm{B} \bm{p_1}^T
\end{equation}

The term \(\bm{D}^{-1} \bm{C}\) in (\ref{eqn_ldpc_enc_p1}) can be pre-computed for a given matrix \(\bm{H}\) using Gaussian elimination and stored in the memory for subsequent iterations. 

\subsection{Baseband Transmitter}

\begin{figure}[!t]
  \centering
  \includegraphics[width=2.6in]{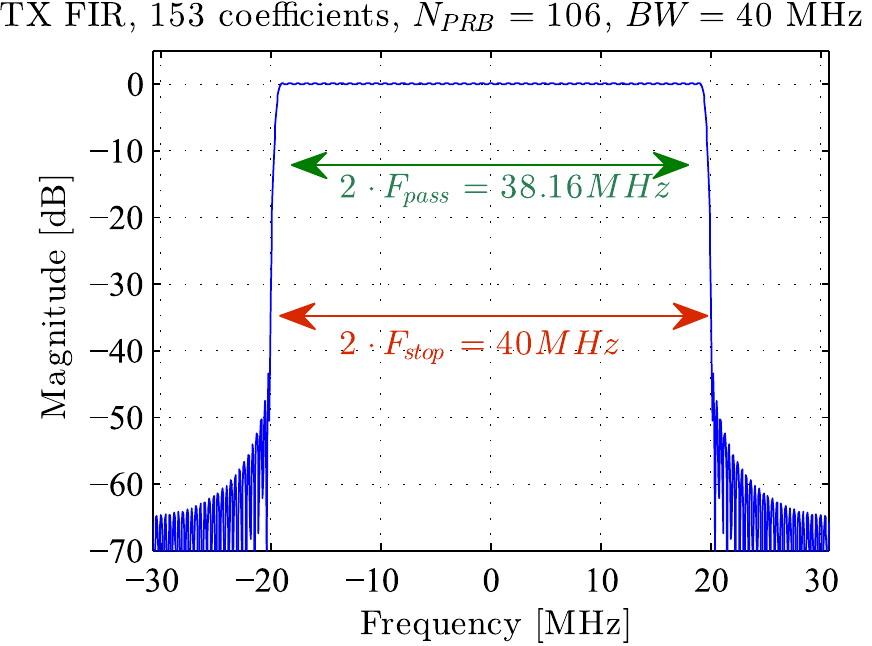}
  \caption{Transmit filter magnitude characteristics.}
  \label{fig_tx_fir}
\end{figure}

The processing performed on physical layer level by the baseband transmitter chain, after LDPC encoding and rate matching, involves scrambling of encoded bits, modulation symbol mapping, DMRS generation and precoding. The procedures are covered in detail by the standard document \cite{TS38211}.

Subcarrier mapped modulation symbols data are OFDMA modulated through IFFT operation. Time domain representations of each symbol are concatenated and filtered using transmit FIR filter to attenuate unwanted Out of Band emission to adjacent frequency bands caused by phase discontinuities and utilization of different numerologies. Filtering may lead to a slight degradation of Error Vector Magnitude (EVM) due to a finite order of the filter \cite{Faulkner00}. The filter is designed to be a real-valued lowpass FIR with passband edge frequency \(F_{\text{pass}}\) and stopband \(F_{\text{stop}}\) given by:
\begin{subequations}
\label{eqn_fir_bands}
\begin{align}
  F_{\text{pass}} &= 0.5 \cdot \Delta f \cdot N_{\text{PRB}} \cdot 12 \\
  F_{\text{stop}} &= 0.5 \cdot \text{BW}
\end{align}
\end{subequations}
where \(\Delta f\) is subcarrier spacing, \(N_{\text{PRB}}\) is a number of PRB and \(\text{BW}\) is the transmission bandwidth. Fig. \ref{fig_tx_fir} shows magnitude characteristics of the filter designed using the Least Squares method.

\subsection{Baseband Receiver}

Consider a two layer transmission with \(M_r = \{0,1\}\) receive and \(M_t = \{0,1\}\) transmit antennas. Once a cyclic prefix removal and OFDM demodulation through IFFT processing have been accomplished, the receiver have to perform an estimation and correction of residual Sample Time Offset (STO) and Carrier Frequency Offset (CFO) \cite{Speth99}. A need for residual synchronization error correction arises from non-ideal adjustment of OFDM FFT window and carrier center frequency done by the UE based on synchronization signal bursts \cite{Omri19}. Both CFO and STO corrections at the receiver have to be performed in frequency domain, because a received signal can be a superposition of transmissions coming from multiple UE multiplexed in frequency, each suffering from a specific residual synchronization error. Residual CFO is estimated as a phase rotation between pilot subcarriers belonging to different OFDM symbols and corrected by a circular convolution operation in frequency domain \cite{Choi00}. Residual STO can be effectively estimated using the DFT method \cite{Kang06}.

Once the synchronization errors are corrected, the next stage incorporates a channel estimation and equalization performed in order to compensate the effects of multipath propagation. A preliminary channel estimate based on Least Squares (LS) criterion is done based on pilot subcarriers. Considering a non-overlapping structure of pilots in spatial domain, the receiver performs a SIMO channel estimation:
\begin{equation}
\label{eqn_chest_ls}
\bm{\hat{h}}_{LS}[m_r,m_t] = \bm{x}^{(p)}[m_t]^H \bm{y}^{(p)}[m_r]
\end{equation}
where \(\bm{x}^{(p)}[m_t]\) and \(\bm{y}^{(p)}[m_r]\) are vectors of pilot symbols transmitted from antenna \(m_t\) and received on \(m_r\). The vector \(\bm{\hat{h}}_{LS}[m_r,m_t]\) is a LS channel estimate for transmission layer \(m_t\) and receiver antenna \(m_r\). Superscript \(H\) denotes a conjugate transpose.

The LS channel estimate is further used to produce an estimate of SNR, which can be derived using a simple and effective estimator proposed by He and Torkelson \cite{He98}, or a more accurate subspace based estimator \cite{Xu04}. The SNR estimate \(\hat{\rho}[m_r,m_t]\) is first calculated per transmission layer and receive antenna, then averaged in order to reduce the overall complexity of the subsequent blocks. The SNR estimation based on \cite{He98} requires calculation of the LS channel estimate \(\bm{\hat{h}}_{LS}\) as given by (\ref{eqn_chest_ls}), which is then de-noised using a moving average filter to produce a smoothed channel estimate \(\bm{\hat{h}}_{LS}^{\text{smooth}}\). Then, the noise estimate is derived as:
\begin{equation}
\label{eqn_chest_snr}
\hat{\rho}[m_r,m_t] = \bigg( \frac{\big\|\bm{\hat{h}}_{LS}^{\text{smooth}}[m_r,m_t]\big\|} {\big\|\bm{\hat{h}}_{LS}^{\text{smooth}}[m_r,m_t] - \bm{\hat{h}}_{LS}[m_r,m_t]\big\|} \bigg) ^ 2
\end{equation}
where operator \(\|\bm{v}\|=\sqrt{\sum_{i=1}^{N} v_i^2 / N}\) denotes Root Mean Square (RMS) of the vector \(\bm{v}=[v_1,...,v_N]\).

The Minimum Mean Squared Error (MMSE) channel estimation is performed in order to minimize an error between the actual and estimated channel realization \cite{Edfors98}:
\begin{equation}
\label{eqn_chest_mmse}
\bm{\hat{h}}_{MMSE}[m_r,m_t] = \bm{R_{hh}} \big(\bm{R_{hh}} + \beta \hat{\rho}^{-1} \bm{I}\big)^{-1} \bm{\hat{h}}_{LS}[m_r,m_t]
\end{equation}
where \(\beta\) is a constant depending on the signal constellation, equal to 1 in case of QPSK. The matrix \(\bm{R_{hh}} = \mathbb{E}\{\bm{h} \bm{h}^H\}\) is a channel autocovariance matrix and \(\bm{I}\) is an identity matrix.

The MMSE estimates of the channel \(\bm{\hat{h}}_{MMSE}[m_r,m_t]\) are then interpolated in time and frequency on non-pilot subcarrier positions. Then, the MMSE equalization is executed on data symbols to reverse the effect of MIMO propagation channel \cite{Mehana12}:
\begin{equation}
\label{eqn_chest_eq}
\bm{\hat{\varsigma}} = \mathcal{G}^H \big(\mathcal{G} \mathcal{G}^H + \hat{\rho}^{-1} \bm{I}\big)^{-1} \bm{\varrho}
\end{equation}
where \(\mathcal{G} = \big[\bm{\hat{h}^{\kappa,\ell}}_{MMSE}[m_r,m_t]\big]_{M_r \times M_t}\) is a MIMO spatial channel matrix for subcarrier at time and frequency positions \(\ell\), \(\kappa\). Analogously, vectors \(\bm{\hat{\varsigma}} = \big[\bm{\hat{x}^{\kappa,\ell}}[m_r]\big]_{M_t \times 1}\) and \(\bm{\hat{\varrho}} = \big[\bm{y^{\kappa,\ell}}[m_r]\big]_{M_r \times 1}\) represent estimated sent and received subcarrier values.

Equalized symbols are demapped and permuted through layer demapping operation to form a one dimensional vector containing all data subcarriers \(\bm{\hat{x}^{\kappa,\ell}}[m_r]\) from all \(M_t\) transmission layers and belonging to the same transport block. Each complex constellation point is demodulated using Maximum \emph{A Posteriori} Probability (MAP) approach to produce values representing beliefs regarding a received bit being 0 or 1, expressed in a form of Log-Likelihood Ratio (LLR). For the sake of the  receiver design, a simplified MAP demodulator proposed by Viterbi \cite{Viterbi98} is used, which approximates logarithmic LLR values of bit \(b\) based on the distance difference between the closest constellation points:
\begin{equation}
\label{eqn_llr_def}
\begin{split}
L(b) & = \log \frac{P(b = 0 | x)}{P(b = 1 | x)} \\
L(b) & \approx - \frac{1}{\sigma^2} \Big( \argmin_{s_0 \in S_0} |x - s_0|^2 \, - \, \argmin_{s_1 \in S_1} |x - s_1|^2 \Big)
\end{split}
\end{equation}
where \(x\) is a received constellation point, \(S_0\) and \(S_1\) are the sets of ideal constellation points corresponding to 0 and 1 respectively,  \(\sigma^2\) is a noise variance of a subcarrier over which the constellation point being demodulated was conveyed.

Soft-demodulated bits are processed using operations specified by 3GPP standard \cite{TS38211, TS38212}, including descrambling, deinterleaving and rate unmatching with LLR soft-combining using circular buffer prior to LDPC decoding. 

\subsection{LDPC Decoder}

Decoding of LDPC codes in practical applications is done based on iterative belief propagation algorithms. An LDPC code can be represented in a form of a bipartite graph with parity check matrix \(\bm{H}\) of size \(M \times N\) being the biadjacency matrix defining connections between the graph nodes: \(M\) rows of the matrix \(\bm{H}\) corresponds to parity check nodes, whereas \(N\) columns corresponds to variable nodes, i.e. the received codeword bits. The principle of the belief propagation algorithms is based on iterative message exchange to update \emph{A Posteriori} probabilities between the variable and the check nodes, until a valid codeword is obtained.

Denote \(\lambda_{n \to m}(d_n)\) as a message sent from variable node \(n\) to check node \(m\) and \(\Lambda_{m \to n}(d_n)\) a message sent from check node \(m\) to variable node \(n\) for a codeword bit \(d_n\), where the codeword is given by \(\bm{d} = [d_0,\, d_1,\, \dots,\, d_{N-1}]\). Moreover, \(\mathcal{N}(m) = \{n \,|\, [\bm{H}]_{m,n} = 1\}\) is a set of variable nodes connected to the check node \(m\) and \(\mathcal{M}(n) = \{m \,|\, [\bm{H}]_{m,n} = 1\}\) is a set of check nodes connected to the variable node \(n\). The process of decoding can be briefly summarized in the following steps \cite{Hu01}:
\begin{itemize}
\item Initialization: assign variable node values to input LLR and check node values to 0 for each \(n\) and \(m\) that satisfy \([\bm{H}]_{m,n} = 1\):
\begin{equation}
\label{eqn_spa_init_v2c}
\lambda_{n \to m} (d_n) = L(d_n)
\end{equation}
\begin{equation}
\label{eqn_spa_init_c2v}
\Lambda_{m \to n} (d_n) = 0
\end{equation}

\item Step I: update the values of check nodes based on the values of variable nodes:
\begin{equation}
\label{eqn_spa_s1_c2v}
\Lambda_{m \to n} (d_n) = 2 \cdot \tanh^{-1} \Big( \displaystyle \prod_{n' \in \mathcal{N}(m) \setminus \{n\}} \tanh \big( \lambda_{n' \to m} (d_{n'}) / 2 \big) \Big)
\end{equation}

\item Step II: update values of variable nodes based on the values of the check nodes from step I and \emph{A Posteriori} LLR of variable nodes:
\begin{equation}
\label{eqn_spa_s2_v2c}
\lambda_{n \to m} (d_n) = L(d_n) + \displaystyle \sum_{m' \in \mathcal{M}(n) \setminus \{m\}} \Lambda_{m' \to n} (d_n)
\end{equation}
\begin{equation}
\label{eqn_spa_s2_v}
\lambda(d_n) = L(d_n) + \displaystyle \sum_{m \in \mathcal{M}(n)} \Lambda_{m \to n} (d_n)
\end{equation}

\item Step III: Perform hard decision on LLR values \(\lambda(d_n)\) to produce codeword \(\bm{\hat{d}} = [\hat{d}_0,\, \hat{d}_1,\, \dots,\, \hat{d}_{N-1}]\), such that \(\hat{d}_n = 0\) when \(\lambda(d_n) > 0\) or \(\hat{d}_n = 1\) otherwise. If the parity check equation \(\bm{H} \bm{\hat{d}} = \bm{0}\) holds, the algorithm terminates.

\end{itemize}

The largest part of LDPC decoder's computational complexity is caused by the check nodes updates performed in step I. In a variant of belief propagation called Sum-Product Algorithm (SPA), the check node updates are computed in pairs of received messages in recursive manner \cite{Hu01, Papaharalabos15}, using a function \(\chi\) called as the Boxplus operator:
\begin{equation}
\label{eqn_spa_rec}
\Lambda_{m \to n} (d_n) = \chi (x_{n-1}, \chi (x_{n-2}, \dots, \chi(x_2, x_1))
\end{equation}
where \(x_i = \lambda_{n' \to m} (d_{n'})\) is the \(i\)-th message connected to the check node \(m\), received from the variable node \(n'\). The Boxplus operator is defined accordingly:
\begin{multline}
\label{eqn_spa_boxplus}
\chi(x_1, x_2) = \text{sign}(x_1) \, \cdot \, \text{sign}(x_2) \, \cdot \, \min(|x_1|, |x_2|) \, + \\ \ln \Big( \frac{1 + \exp (-|x_1+x_2|)}{1 + \exp (-|x_1-x_2|)} \Big)
\end{multline}

For the sake of the presented decoder implementation, a low cost approximation of the term \(\ln (1 + e^{-|x|} )\) in (\ref{eqn_spa_boxplus}) by a two piece linear function is used, as proposed by Richter \emph{et. al.} \cite{Richter05}.
\begin{equation}
\label{eqn_spa_boxplus_twopiece}
\ln (1 + e^{-|x|} ) \approx 
  \begin{cases}
    0.6 - 0.24 \cdot |x| & \text{if} |x| < 2.5 \\
    0 & \text{otherwise}
  \end{cases}
\end{equation}

It was confirmed in \cite{Butler12} that the approximation (\ref{eqn_spa_boxplus_twopiece}) provides several time complexity reduction with negligible performance degradation in AWGN channel.

\section{Implementation Performance}

Performance of the PUSCH software transceiver implementation was evaluated in term of Block Error Ratio (BLER) metric, defined as a ratio between the number of received code block with failed CRC check to the total number of transmitted blocks. The simulations were conducted considering the system parameters listed in Table \ref{table_testcase}, for selected Modulation and Coding Schemes (MCS) listed in Table \ref{table_mcs}. 

\begin{figure}[!t]
  \centering
  \includegraphics[width=3.5in]{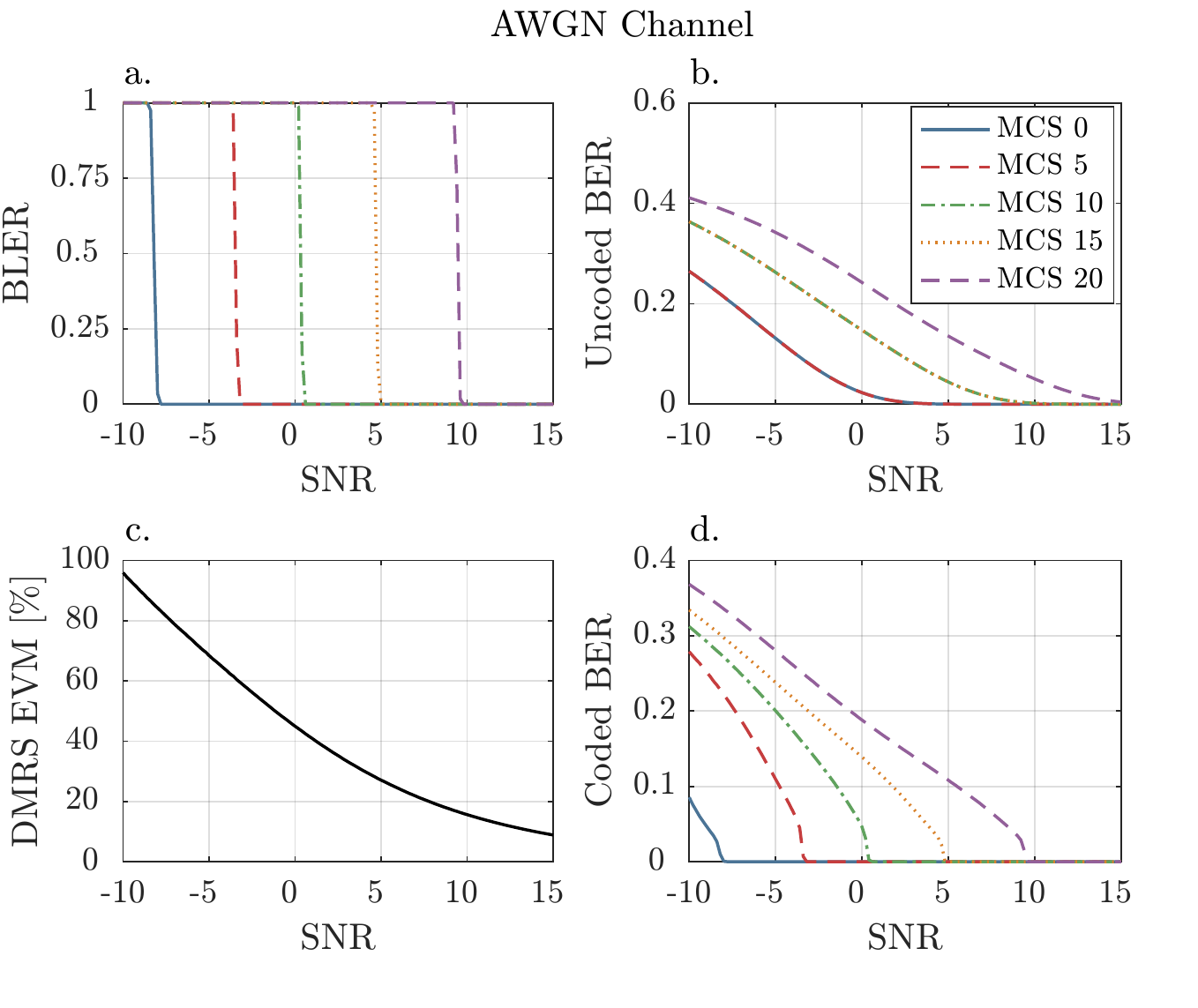}
  \caption{Performance of the system in AWGN channel under various MCS configurations and SNR: a) Block Error Ratio (BLER); b) Bit Error Ratio (BER) of demodulated IQ symbols before decoding; c) Error Vector Magnitude (EVM) of equalized DMRS signal; d) BER of information bits after decoding.}
  \label{fig_bler_awgn}
\end{figure}

Results of the simulations for pure AWGN channel are depicted in Fig. \ref{fig_bler_awgn}. In this scenario, no multipath fading or user mobility are taken into considerations, so the focus is set on the performance of LDPC decoder part. Obtained curves shows a very sharp \emph{waterfall} decrease of BLER below some SNR threshold for all considered MCS. A similar results were reported by other 5G NR prototype implementations \cite{Ji18}. Such phenomenon characterizes a well designed communication system based on FEC.

\begin{figure}[!t]
  \centering
  \includegraphics[width=3.5in]{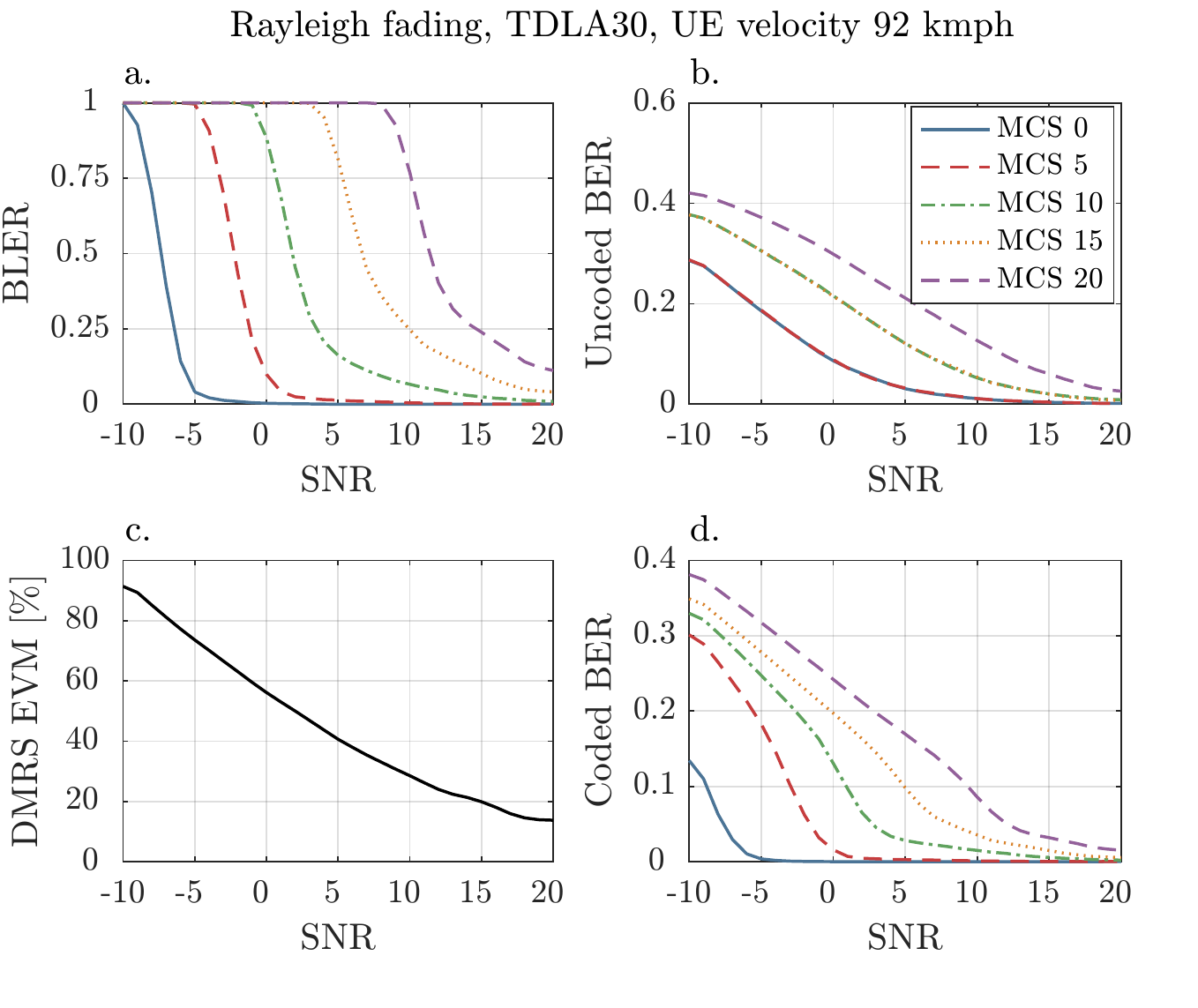}
  \caption{Performance of the system under Rayleigh fading. The channel considers multipath propagation model TDLA30 \cite{TS38104} with the maximum Doppler frequency of 300 Hz.}
  \label{fig_bler_tdl}
\end{figure}

Fig. \ref{fig_bler_tdl} shown the results considering Rayleigh fading, multipath and MIMO propagation characteristics as specified by the Annex G of \cite{TS38104}. The Rayleigh fading distribution of multipath channel taps is obtained using the Sum-of-Sinusoids model \cite{Zheng02}. The steepness of the slope of BLER curves is lower than in case of AWGN channel. The performance decrease of the system in case of higher modulation orders is a consequence of several factors, including presence of Intercarrier Interference due to user mobility not compensated by the receiver, finite equalizer precision due to channel matrix singularity, non-ideal channel estimate interpolation and variance of parameter estimators.

\setlength\tabcolsep{7pt}
\begin{table}[!t]
\renewcommand{\arraystretch}{1.2}
\caption{Simulation parameters.}
\label{table_testcase}
\centering
\begin{tabular}{|c|c|}
\hline
Carrier frequency                & 3.5 GHz (band n78) \\
\hline
Transmission mode                & 2x2 Spatial Multiplexing  \\
\hline
Antenna configuration            & Uniform Linear Array \\
\hline
Antenna correlation              & Low \\
\hline
Bandwidth                        & 40 MHz  \\
\hline
Subcarrier spacing \(\Delta f\)  & 30 kHz  \\
\hline
FFT size \(N_{\text{FFT}}\)      & 2048 \\
\hline
Number of PRB \(N_{\text{PRB}}\) & 106 \\
\hline
Operating subcarriers            & 1272 \\
\hline
Cyclic prefix length \(N_{\text{CP}}\)  & 176 long, 144 short \\
\hline
Number of data symbols in slot   & 11 \\
\hline
Number of DMRS symbols in slot   & 2 \\
\hline
DMRS separation in frequency     & 2 subcarriers \\
\hline
PTRS configuration               & Disabled \\
\hline
Transform precoding              & Disabled \\
\hline
Channel estimator                & MMSE \\
\hline
Channel estimation interpolation & Cubic Spline \\
\hline
Equalizer                        & MMSE \\
\hline
LDPC decoding algorithm          & Sum Product Algorithm (SPA) \\
\hline
Transmit filter coefficients     & 153 \\
\hline
\end{tabular}
\end{table}

\setlength\tabcolsep{12pt}
\begin{table}[!t]
\renewcommand{\arraystretch}{1.2}
\caption{MCS configurations used for evaluation of the system implementation. Transport Block Sizes (TBS) were obtained based on the configuration listed in Table \ref{table_testcase}.}
\label{table_mcs}
\centering
\begin{tabular}{cccc}
\hline
MCS Index  & Modulation  & Code Rate & TBS \\
\hline
0    & QPSK        & 0.117 & 7176   \\
5    & QPSK        & 0.370 & 22536  \\
10   & 16-QAM      & 0.332 & 40976  \\
15   & 16-QAM      & 0.602 & 73776  \\
20   & 64-QAM      & 0.554 & 102416 \\
\hline
\end{tabular}
\end{table}

\section{Conclusion}

A prototype implementation of the 5G NR PUSCH transceiver has been reported in the paper. It was shown that considerable BLER performance can be achieved by using a well-founded algorithms developed for legacy 3G and 4G systems. The structure of PUSCH channel in 5G NR has much greater flexibility comparing to 4G LTE, but the overall transceiver structure remains similar with minor modifications, except for the pilot symbol arrangements and FEC.

The authors believe that the outcome of the reported work may be used as a reference for other software or hardware implementations of the 5G NR physical layer. Moreover, the existing model may be further extended by adding a support for the remaining 5G NR physical channels and procedures to form a 5G NR link-level simulator.

\appendices

\section{Code Repository}

The source code of the high-level software transceiver numerical model written in MATLAB is available online on a public repository under an open source license. The model has been proven to be compatible with GNU Octave version 4.2.0 and MATLAB version 2013B. The repository can be accessed under the following URL:

\texttt{\url{https://github.com/gc1905/5g-nr-pusch}}

\section*{Acknowledgment}

This work was supported by the Polish Ministry of Science and Higher Education with the subvention funds of the Faculty of Computer Science, Electronics and Telecommunications of AGH University (contract no 16.16.230.434).

%The work was supported by AGH University of Science and Technology contract no 16.16.230.434.

\balance


\begin{thebibliography}{1}

\bibitem{Hou13}
N.~Hou \emph{et. al.}, "Test and Performance Analysis of PUSCH Channel of LTE System," in \emph{5th IEEE Int. Symp. on Microwave, Antenna, Propagation and EMC Technol. for Wireless Commun.}, Oct. 2013.

\bibitem{Levanen18}
T.~Levanen \emph{et. al.}, "5G New Radio Uplink Performance: Noise, Interference and Emission Constraints," in \emph{IEEE Wireless Commun. and Netw. Conf. (WCNC)}, April 2018.

\bibitem{DiStasio18}
F.~Di Stasio, M.~Mondin and F.~Daneshgaran, "Multirate 5G Downlink Performance Comparison for f-OFDM and w-OFDM Schemes with Different Numerologies," in \emph{Int. Symp. on Netw., Computers and Commun. (ISNCC)}, June 2018.

\bibitem{Richardson18}
T.~Richardson and S.~Kudekar, "Design of Low-Density Parity Check Codes for 5G New Radio," \emph{IEEE Commun. Mag.}, vol. 56, no. 3, pp. 28-34, March 2018.

\bibitem{Shao19}
S.~Shao \emph{et. al.}, "Survey of Turbo, LDPC and Polar Decoder ASIC Implementations," \emph{IEEE Commun. Surveys Tuts.}, Early Access, Jan. 2019.

\bibitem{TS38211}
3GPP TS 38.211, "NR; Physical Channels and Modulation," version 15.0.0, December 2017.

\bibitem{TS38212}
3GPP TS 38.212, "NR; Multiplexing and Channel Coding," version 15.0.0, December 2017.

\bibitem{Zaidi18}
A.~Zaidi \emph{et. al.}, "OFDM Numerology Design for 5G New Radio to Support IoT, eMBB, and MBSFN," \emph{IEEE Commun. Standards Mag.}, vol. 2, no. 2, pp. 78 - 83, June 2018.

\bibitem{Noh17}
G.~Noh \emph{et. al.}, "DMRS Design and Evaluation for 3GPP 5G New Radio in a High Speed Train Scenario," in \emph{IEEE Global Commun. Conf. (GlobeCom)}, Dec. 2017.

\bibitem{Qi18}
Y.~Qi \emph{et. al.}, "On the Phase Tracking Reference Signal (PT-RS) Design for 5G New Radio (NR)," in \emph{IEEE 88th Veh. Technol. Conf. (VTC-Fall)}, Aug. 2018.

\bibitem{Chu08}
C.~Chu, C.~Lee and Y.~Huang, "Design of an OFDMA Baseband Receiver for 3GPP Long-Term Evolution," in \emph{IEEE Int. Symp. VLSI Design, Automation and Test (VLSI-DAT)}, April 2008.

\bibitem{Hamidi18}
F.~Hamidi-Sepehr, A.~Nimbalker and G.~Ermolaev, "Analysis of 5G LDPC Codes Rate-Matching Design," in \emph{IEEE 87th Veh. Technol. Conf. (VTC-Spring)}, June 2018.

\bibitem{Richardson01}
T.~Richardson and R.~Urbanke, "Efficient Encoding of Low-Density Parity-Check Codes," \emph{IEEE Trans. Inf. Theory}, vol. 47, no. 2, pp. 638-656, Feb. 2001.

\bibitem{Faulkner00}
M.~Faulkner, "The Effect of Filtering on the Performance of OFDM Systems," \emph{IEEE Trans. Veh. Technol.}, vol. 49, no. 5, pp. 1877-1884, Sept. 2000.

\bibitem{Speth99}
M.~Speth \emph{et. al.}, "Optimum Receiver Design for Wireless Broad-Band Systems Using OFDM—Part I," \emph{IEEE Trans. Commun.}, vol. 47, no. 1, pp. 1668-1677, Nov. 1999.

\bibitem{Omri19}
A.~Omri \emph{et. al.}, "Synchronization Procedure in 5G NR Systems," \emph{IEEE Access}, vol. 7,  pp. 41286-41295, March 2019.

\bibitem{Choi00}
J.~Choi \emph{et. al.}, "Carrier Frequency Offset Compensation for Uplink of OFDM-FDMA Systems", \emph{IEEE Commun. Let.}, vol. 4, no. 12, Dec. 2000.

\bibitem{Kang06}
Y.~Kang, D.~Ahn and H.~Lee, "OFDM Channel Estimation with Timing Offset for Satellite plus Terrestrial Multipath Channels," in \emph{IEEE 63rd Veh. Technol. Conf.}, May 2006.

\bibitem{He98}
S.~He and M.~Torkelson, "Effective SNR Estimation in OFDM System Simulation," in \emph{IEEE Global Commun. Conf. (GLOBECOM)}, Nov. 1998.

\bibitem{Xu04}
X.~Xu, Y.~Jing and X.~Yu, "Subspace-Based Noise Variance and SNR Estimation for OFDM Systems," in \emph{IEEE Wireless Commun. Netw. Conf.}, March 2005.

\bibitem{Edfors98}
O.~Edfors \emph{et. al.}, "OFDM Channel Estimation by Singular Value Decomposition," \emph{IEEE Trans. Commun.}, vol. 46, no. 7, pp. 931-939, July 1998.

\bibitem{Mehana12}
A.~Mehana and A.~Nosratinia, "Diversity of MMSE MIMO Receivers," \emph{IEEE Trans. Inf. Theory}, vol. 58, no. 11, pp. 6788-6805, Nov. 2012.

\bibitem{Viterbi98}
A.~Viterbi, "An Intuitive Justification and a Simplified Implementation of the MAP Decoder for Convolutional Codes," \emph{IEEE J. Sel. Areas Commun.}, vol. 16, no. 2, pp. 260-264, Feb. 1998.

\bibitem{Hu01}
X.~Hu \emph{et. al.}, "Efficient Implementations of the Sum-Product Algorithm for Decoding LDPC Codes" in \emph{IEEE Global Commun. Conf. (GLOBECOM)}, Nov. 2001. 

\bibitem{Papaharalabos15}
S.~Papaharalabos and F.~Lazarakis, "Approximated Box-Plus Decoding of LDPC Codes," \emph{IEEE Commun. Lett.}, vol. 19, no. 12, pp. 2074-2077, Dec. 2015.

\bibitem{Richter05}
G.~Richter \emph{et. al.},  “Optimization of a Reduced-complexity Decoding Algorithm for LDPC Codes by Density Evolution,” in \emph{IEEE Int. Conf. Commun.}, May 2005.

\bibitem{Butler12}
B.~Butler and P.~Siegel, "Numerical Issues Affecting LDPC Error Floors," in \emph{IEEE Global Commun. Conf. (GLOBECOM)}, Dec. 2012.

\bibitem{Ji18}
W.~Ji \emph{et. al.}, "Design and Implementation of a 5G NR System Based on LDPC in Open Source SDR," in \emph{IEEE Globecom Workshops (GC Wkshps)}, Dec. 2018.

\bibitem{TS38104}
3GPP TS 38.104, "NR; Base Station (BS) Radio Transmission and Reception," version 15.4.0, Dec. 2018.

\bibitem{Zheng02}
Y.~Zheng and C.~Xiao, "Improved Models for the Generation of Multiple Uncorrelated Rayleigh Fading Waveforms," \emph{IEEE Commun. Lett.}, vol. 6, no. 6, pp. 256–258, June 2002.

%\bibitem{Zivkovic17} % NOMA in 5G
%M.~Zivkovic \emph{et. al.}, "An SDR-based Turbo-SIC Implementation: Towards a 5G New Radio Advanced Receiver for Uplink Boosting," in \emph{IEEE 28th Annual Int. Symp. Personal, Indoor and Mobile Radio Commun. (PIMRC)}, Oct. 2017.

% On the Performance of PDCCH in LTE and 5G New Radio
% 5G NR PDCCH: Design and Performance  

% Physical layer performance for low latency and high reliability in 5G
% URLLC in 5G

% Filtered OFDM: A New Waveform for Future Wireless Systems

\end{thebibliography}
\end{document}